\documentclass[]{spie}  

\newcommand{\oiii}{[\textsc{O\,iii}]}
\newcommand{\cii}{[\textsc{C\,ii}]}

\usepackage{amsmath,amsfonts,amssymb}
\usepackage{graphicx}
\usepackage[colorlinks=true, allcolors=blue]{hyperref}
\usepackage{multirow}

\title{FINER: Far-Infrared Nebular Emission Receiver for the Large Millimeter Telescope}

\author[a]{Yoichi Tamura}
\author[b]{Takeshi Sakai}
\author[c]{Ryohei Kawabe}
\author[c]{Takafumi Kojima}
\author[a]{Akio Taniguchi}
\author[d]{Tatsuya Takekoshi}
\author[c]{Haoran Kang}
\author[c]{Wenlei Shan}
\author[a]{Masato Hagimoto}
\author[a]{Norika Okauchi}
\author[b]{Airi Tetsuka}
\author[e]{Akio K.~Inoue}
\author[f]{Kotaro Kohno}
\author[g]{Kunihiko Tanaka}
\author[h]{Tom J.L.C.~Bakx}
\author[i]{Yoshinobu Fudamoto}
\author[j]{Kazuyuki Fujita}
\author[k]{Yuichi Harikane}
\author[l]{Takuya Hashimoto}
\author[c]{Bunyo Hatsukade}
\author[m]{David H.~Hughes}
\author[n]{Takahiro Iino}
\author[j]{Yuki Kimura}
\author[o]{Hiroyuki Maezawa}
\author[c]{Yuichi Matsuda}
\author[e]{Ken Mawatari}
\author[p]{Taku Nakajima}
\author[q]{Shunichi Nakatsubo}
\author[c]{Tai Oshima}
\author[r]{Hideo Sagawa}
\author[s]{F.~Peter Schloerb}
\author[n]{Shigeru Takahashi}
\author[c]{Kotomi Taniguchi}
\author[f]{Akiyoshi Tsujita}
\author[a]{Hideki Umehata}
\author[o]{Teppei Yonetsu}
\author[s]{Min S.~Yun}

\affil[a]{Department of Physics, Nagoya University, Nagoya, Aichi 464-8602, Japan}
\affil[b]{University of Electro-Communications, Chofu, Tokyo 182-8585, Japan}
\affil[c]{National Astronomical Observatory of Japan, Mitaka, Tokyo 181-8588, Japan}
\affil[d]{Kitami Institute of Technology, Kitami, Hokkaido 090-8507, Japan}
\affil[e]{Waseda University, Shinjuku, Tokyo 169-8555, Japan}
\affil[f]{Institute of Astronomy, The University of Tokyo, Mitaka, Tokyo 181-0015, Japan}
\affil[g]{Keio University, Yokohama, Kanagawa 223-8522, Japan}
\affil[h]{Chalmers University of Technology, Onsala Observatory, SE-439 94 Onsala, Sweden}
\affil[i]{Chiba University, Chiba 263-8522, Japan}
\affil[j]{Institute of Low Temperature Science, Hokkaido University, Sapporo, Hokkaido 060-0819, Japan}
\affil[k]{Institute of Cosmic Ray Research, The University of Tokyo, Kashiwa, Chiba 277-8582, Japan}
\affil[l]{Tsukuba University, Tsukuba, Ibaraki 305-8751, Japan}
\affil[m]{Instituto Nacional de Astrof\'{i}sica, \'{O}ptica y Electr\'{o}nica (INAOE), Tonantzintla, Puebla 72480, Mexico}
\affil[n]{Information Technology Center, The University of Tokyo, Kashiwa, Chiba 277-0882, Japan}
\affil[o]{Osaka Metropolitan University, Sakai, Osaka 599-8531, Japan}
\affil[p]{Institute for Space-Earth Environmental Research, Nagoya University, Nagoya, Aichi 464-8601, Japan}
\affil[q]{Institute of Space and Astronautical Science, Japan Aerospace Exploration Agency, Sagamihara, Kanagawa 252-5210, Japan}
\affil[r]{Kyoto Sangyo University, Kyoto 603-8555, Japan}
\affil[s]{University of Massachusetts, Amherst, MA 01003, USA}

\authorinfo{Further author information: (Send correspondence to Y.T.)\\Y.T.: E-mail: ytamura@nagoya-u.jp}

\begin{document} 
\maketitle

\begin{abstract}
Unveiling the emergence and prevalence of massive/bright galaxies during the epoch of reionization and beyond, within the first 600 million years of the Universe, stands as a pivotal pursuit in astronomy.  Remarkable progress has been made by \textit{JWST} in identifying an immense population of bright galaxies, which hints at exceptionally efficient galaxy assembly processes. However, the underlying physical mechanisms propelling their rapid growth remain unclear. With this in mind, millimeter and submillimeter-wave spectroscopic observations of redshifted far-infrared spectral lines, particularly the \oiii{} 88~$\mu$m and \cii{} 158~$\mu$m lines, offers a crucial pathway to address this fundamental query.

To this end, we develop a dual-polarization sideband-separating superconductor-insulator-superconductor (SIS) mixer receiver, FINER, for the Large Millimeter Telescope (LMT) situated in Mexico. Harnessing advancements from ALMA's wideband sensitivity upgrade (WSU) technology, FINER covers radio frequencies spanning 120--360~GHz, delivering an instantaneous intermediate frequency (IF) of 3--21 GHz per sideband per polarization, which is followed by a set of 10.24 GHz-wide digital spectrometers. At 40\% of ALMA's light-collecting area, the LMT's similar atmospheric transmittance and FINER's 5 times wider bandwidth compared to ALMA culminate in an unparalleled spectral scanning capability in the northern hemisphere, paving the way for finer spectral-resolution detection of distant galaxies. 
\end{abstract}

\keywords{Millimeter and submillimeter instrumentation, Heterodyne receiver, SIS receiver, LMT, ALMA, High-redshift universe, Galaxy formation, Interstellar medium}

\section{INTRODUCTION}
\label{sec:intro}  

\subsection{Scientific Background and Key Science}
\begin{figure}
    \centering
    \includegraphics[width=0.95\linewidth]{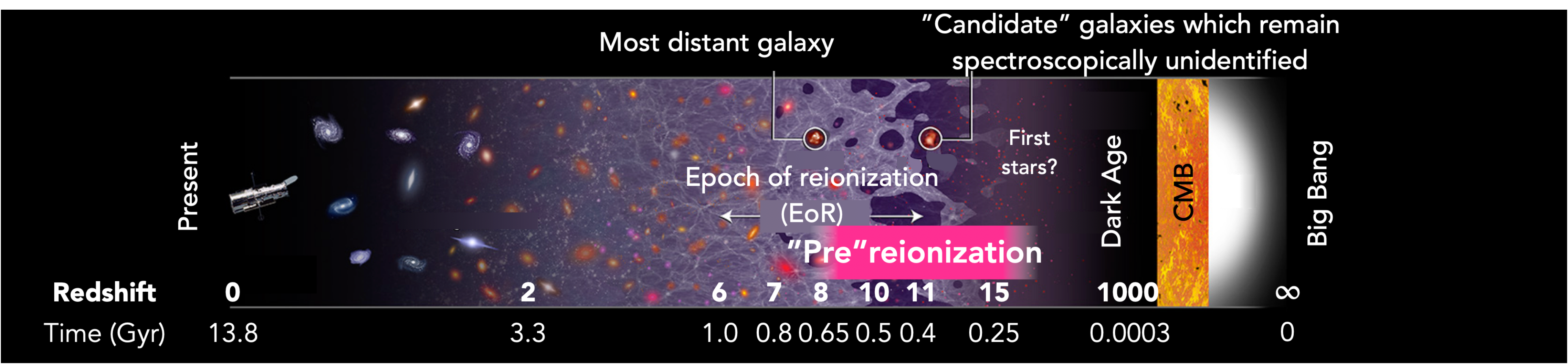}
    \caption{Galaxy formation at the epoch of reionization and beyond (pre-EoR) in the history of the Universe. [Image credit: NASA, ESA, P.\ Oesch and B.\ Robertson (University of California, Santa Cruz), and A.\ Feild (STScI)]}
    \label{fig:1}
\end{figure}

How and when first stars and galaxies form is one of the most fundamental questions in modern astronomy. The pristine materials in the early universe cooled down to form neutral hydrogen at redshift $z \sim 1000$ or 380k years after the Big Bang. Once first stars and galaxies are born, however, the hydrogen gas in the intergalactic space is `reionized' by hydrogen-ionizing photons from the hot, massive stars. This so-called cosmic reionization is indirect evidence for the birth of the first luminous objects. In recent years, Planck satellite has reported that cosmic reionization occurred in the period centered at redshift $z = 7.8$ (660 million years after the Big Bang),\cite{Planck20} whereas hundreds of galaxy candidates have been identified with \textit{Hubble} and \textit{James Webb Space Telescopes} (\textit{HST} and \textit{JWST}) even before the period\cite{Castellano22, Castellano23, Finkelstein22, Naidu22, Harikane23, Harikane24, Atek23, Donnan23, Bouwens23}, meaning that we are likely witnessing the birth of the earliest galaxies in the universe (Figure~\ref{fig:1}).

In particular, finding massive galaxies in the era is difficult but is expected to offer unique information on galaxy formation, because the probability of finding them (i.e., volume density or luminosity function) depends on cosmic structure formation models. Recent discoveries of the excessive massive galaxy candidates at $z > 8$ made by \textit{HST} and \textit{JWST} suggest an extremely high efficiency of galaxy assembly in the earliest Universe, while it is still unclear what drives the rapid growth of the massive population. Then, the key questions we are trying to address in this research include 
(i) when massive galaxies emerge, 
(ii) how common they are in the pre-reionization era, and
(iii) what controls their growth.
It is, however, difficult to confirm galaxy `candidates' in the era; in general, identifying an object as a galaxy and determining its physical properties begin with spectroscopy of atomic or molecular spectral lines. But, the combination of bright spectral lines and a sensitive instrument that are sufficient for detection of a galaxy even in the furthest universe was very limited.

\begin{figure}
    \centering
    \includegraphics[width=0.80\linewidth]{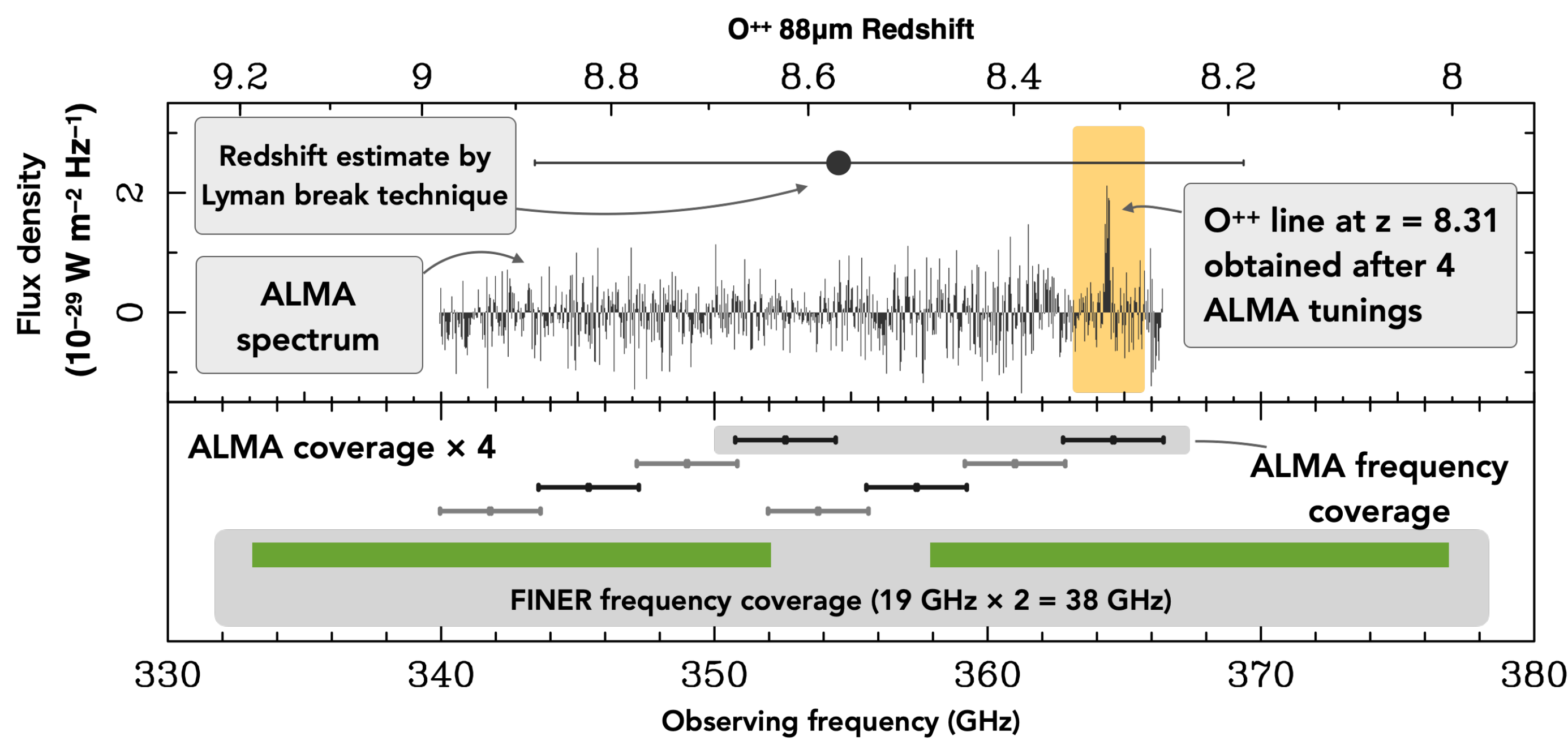}
    \caption{The \oiii{} 88 $\mu$m line detected a $z = 8.31$ Lyman break galaxy, MACS0416\_Y1\cite{Tamura19}. The spectrum was obtained by four tunings of ALMA. Redshift uncertainty of a typical galaxy candidate at $z \sim 10$ is $\Delta z \sim 1$, requiring a $> 30$~GHz spectral coverage. ALMA requires 4 to 6 tunings, whereas FINER only needs a single tuning to cover $> 30$~GHz.}
    \label{fig:3}
\end{figure}

In this study, we aim for a sensitive spectroscopic survey of galaxies in the pre-reionization era using the far-infrared fine-structure lines, the $^3\!P_1 \rightarrow\, ^3\!P_0$ transition of doubly-ionized oxygen \oiii{} 88.3~$\mu$m (3393.01~GHz) and the $^2\!P_{1/2}\rightarrow\, ^2\!P_{3/2}$ transition of ionized carbon \cii{} 157.7~$\mu$m (1900.54~GHz), based on our novel mm and submm-wave receiver, FINER, for the Large Millimeter Telescope (LMT).\cite{Hughes10, Hughes20} The \oiii{} line is demonstrated to be bright in ultraviolet-selected galaxies\cite{Inoue16} and now becomes a useful tracer in identifying their spectroscopic redshifts\cite{Hashimoto18, Tamura19} (Figure~\ref{fig:3}) even compared with \textit{JWST}. We also aim to elucidate their physical properties, such as gas-phase metallicity, ionizing photon and gas densities, and multi-phase gas structure, such as a covering fraction of neutral gas in photo-dissociation regions. Those physics are the key to unveil the origins of the extraordinary efficiency of star formation in the earliest massive galaxies \textit{JWST} has found.

\subsection{Broad Impact and Complementary Role of FINER}
\begin{figure}
    \centering
    \includegraphics[width=0.90\linewidth]{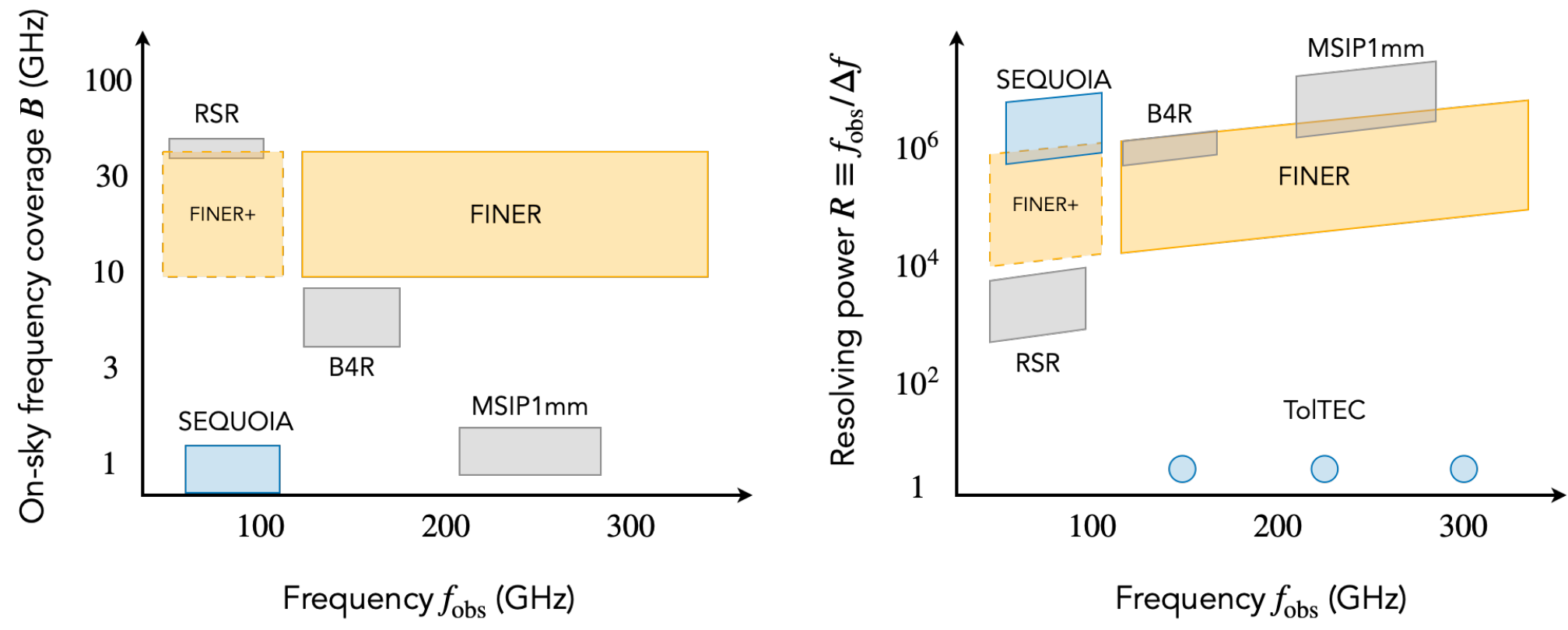}
    \caption{Instantaneous on-sky frequency coverage and resolving power of LMT receivers as a function of frequency. The gray and blue rectangles represent the capabilities of existing single and multi-beam receivers/camera, respectively. FINER and its potential 3-mm extension (denoted as FINER+, see also Section~\ref{sect:DevPlan}) are represented by the orange rectangles.}
    \label{fig:5}
\end{figure}

With these capabilities, FINER offers broader science cases in addition to the EoR galaxy science (see Table~\ref{tab:1}), such as solar system, interstellar medium and astrochemistry, star formation, nearby and distant submillimeter-bright galaxies. 
Figure~\ref{fig:5} shows how FINER complements the capabilities of LMT's existing receivers. FINER fully covers the same frequency bands as TolTEC but with broad, high-dispersion spectroscopic capabilities. Although FINER is a single-beam receiver, it offers the imaging spectroscopic capability, which complements SEQUOIA and allows one to seamlessly cover the mm to submm bands. As shown in the right panel of Figure~\ref{fig:5}, FINER is the unique instrument offering the highest spectral-resolving power $R \equiv f/\Delta f$ over the frequency range from 125 to 210 GHz, which is useful for planetary, Galactic, and extragalactic spectroscopic studies and thus will benefit the LMT user community. It also offers the widest instantaneous frequency coverage with a modest spectral resolving power $R = 10,500$--17,500 among the LMT receivers operating at 120--360 GHz (the bottom edge of the `FINER' rectangle in Figure~\ref{fig:5} left).\footnote{New capabilities will be delivered by the tri-band receiver for Event Horizon Telescope and a 1-mm array receiver OMAyA, which are being developed.\cite{Hughes20}}

In addition, the improved continuum sensitivity offered by FINER's wide frequency coverage will benefit telescope calibration measurements, such as relative pointing, focusing, astigmatism and beam efficiency measurements. The instantaneous IF bandwidth covered by the DRS4 spectrometers is 33~GHz per polarization, which is approximately $20\times$ wider than the current per-polarization bandwidth of MSIP1mm. This wideband capability will increase the continuum sensitivity by a factor of $\approx 4.5$, contributing not only to the improvement in the calibration accuracy but also to a substantial decrease in the overhead time for telescope calibration. Accurate measurements of telescope performances with a single receiver system will reduce potential systematics of the measurements and will allow us to more precisely understand the thermal, mechanical, and optical performances of the telescope across a wide frequency range from 120 to 360~GHz. This can also be verified and cross-checked by TolTEC, providing reliable results for the telescope parameters that benefit every type of science observation with the LMT.

\begin{table}[ht]
\begin{center}       
\caption{Summary of potential science cases afforded by LMT-FINER.} 
\label{tab:1}
\begin{tabular}{l|l} 
\hline
    \rule[-1ex]{0pt}{3.5ex}  \bf Scientific field & \bf Use cases  \\
\hline
    \rule[-1ex]{0pt}{3.5ex}  Solar system 
        &  $\bullet$ High resolution spectral-line mapping of planetary atmosphere \\
        &  $\bullet$ Molecular line survey \\
        &  $\bullet$ ToO/monitoring observations \\
\hline
    \rule[-1ex]{0pt}{3.5ex}  Star-formation and astrochemistry 
        &  $\bullet$ Molecular line survey \\
        &  $\bullet$ High angular resolution spectral-line mapping \\
\hline
    \rule[-1ex]{0pt}{3.5ex}  Dusty star forming galaxies 
        &  $\bullet$ Redshift identification in CO, [CI], and [CII] \\
        &  $\bullet$ CO spectral energy distribution \\
        &  $\bullet$ Physical and chemical properties of ISM \\
\hline
    \rule[-1ex]{0pt}{3.5ex}  Galaxy formation 
        &  $\bullet$ Redshift identification in [OIII] and [CII] \\
        &  $\bullet$ Physical and chemical diagnostics of ISM \\
\hline
\end{tabular}
\end{center}
\end{table} 

\section{INSTRUMENT}

FINER is a dual-polarization sideband-separating superconductor-insulator-superconductor (SIS) heterodyne receiver system which comprises two frequency bands covering 120--210~GHz (Band 4+5) and 210--360~GHz (Band 6+7), each of which has an intermediate frequency (IF) bandwidth of 3--21~GHz per sideband. The feeds of the two frequency bands are placed on the focal plane separately, allowing simultaneous dual-band observations with an on-sky beam separation of 60~arcsec. FINER employs novel SIS mixers with a high critical current density ($J_{\rm c}$) circuit to increase the IF bandwidth\cite{Kojima17, Kojima20}. It is also equipped with an array of FPGA-based digital spectrometers with the total bandwidth of 163.84~GHz ($16 \times 10.24$~GHz).

\begin{figure}
    \centering
    \includegraphics[width=0.6\linewidth]{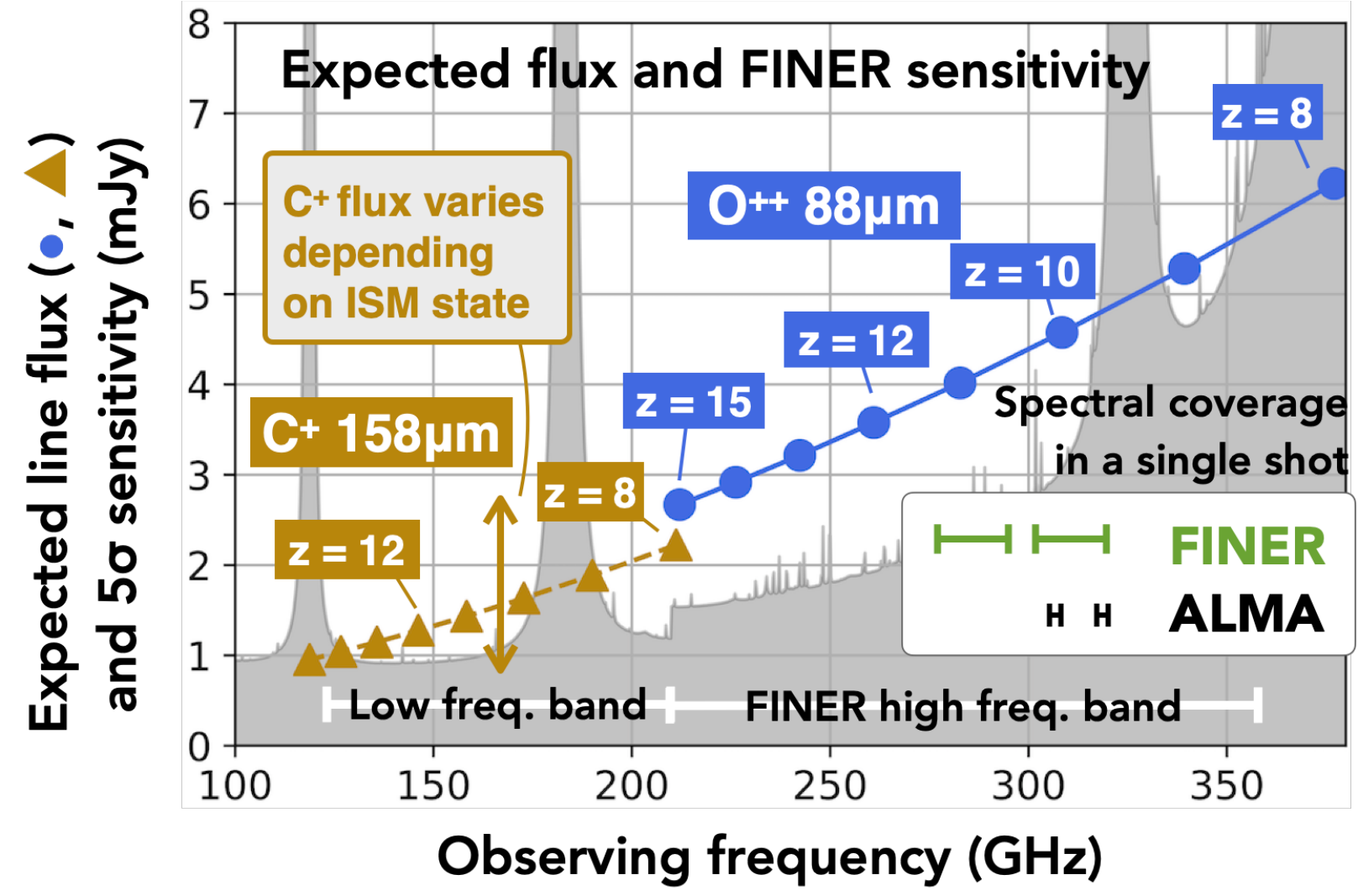}
    \caption{The expected intensities of the \oiii{} and \cii{} lines expected for a bright Lyman break galaxy with an apparent rest-UV magnitude of $m_{\rm AB} = 25$ (blue and yellow curves) and the $5\sigma$ limiting sensitivity of LMT-FINER (gray shade) as a function of observing frequency. The sensitivity is estimated for a 10-hour on-source observation.}
    \label{fig:10}
\end{figure}

This offers an instantaneous bandwidth $5\times$ wider than ALMA, while achieving the same sensitivity as ALMA (Figure~\ref{fig:3}). LMT-FINER will provide 40\% of ALMA's light-collecting area, a similar atmospheric transmittance to ALMA, and a $5\times$ wider bandwidth than ALMA. This offers the most sensitive spectral-scanning capability among mm/submm telescopes in the northern hemisphere (Figure~\ref{fig:10}).

\subsection{Science Requirements and Specifications}
\label{sec:spec}

The level-1 science requirements are defined as follows.
\begin{itemize}
    \item A single Lyman break galaxy with $H = 25.0$ AB and $\Delta z/(1+z) = 0.1$ needs to be detected in \oiii{} 88~$\mu$m at a S/N = 5 or higher within a single night (8~hr) observation. 
    \item The redshifted \oiii{} and \cii at $z = 9$ to 15 need to be observed by a tunable receiver. 
    \item The expected peak flux density of \oiii{} 88~$\mu$m is 5~mJy at $z = 10$.
\end{itemize}
The level-1 requirements are broken down to the technical specifications as listed in the science traceability matrix (Table \ref{tab:2}). Table~\ref{tab:3} summarizes the specifications of FINER. Figure~\ref{fig:10} shows how the specified capabilities meet the science requirements.

\begin{table}
\centering
\caption{Science traceability matrix.}
\label{tab:2}
\begin{tabular}{p{7em}|p{7.5em}|l}
\hline
    \bf Level & \multicolumn{2}{l}{\bf Requirements} \\
\hline

\begin{tabular}{l}Level 1:\\ Science\\ objectives\end{tabular}
    &  \multicolumn{2}{l}{
        \begin{tabular}{l}$
            \bullet$ To measure the bright-end UVLF ($M_{\rm UV} < -22$) of galaxies at $z = 8.7$ to 11. \\
            $\bullet$ To constrain the bright-end UVLF ($M_{\rm UV} < -22.5$) of galaxies at $z = 12$ to 15. \\
            $\bullet$ To constrain a covering fraction of neutral medium surrounding the ionized regions\\ of bright-end galaxies.
        \end{tabular}}\\

\hline

\multirow{7}{*}{\begin{tabular}{l}Level 2:\\ Science\\ requirements\end{tabular}}
    &  \begin{tabular}{l}Science\\ observables\end{tabular} 
        &  \begin{tabular}{l}
            $\bullet$ Spectroscopic redshift of a candidate galaxy selected by the\\ Lyman break technique. \\
            $\bullet$ \cii{} 158 $\mu$m to \oiii{} 88 $\mu$m luminosity ratio. \\
            $\bullet$ The number of spectroscopically-identified galaxies.
        \end{tabular} \\
        
    \cline{2-3}
    
    &  \begin{tabular}{l}Measurement\\ requirements\end{tabular}
        &  \begin{tabular}{l}
            $\bullet$ To detect \oiii{} 88 $\mu$m of a candidate $z \sim 10$ galaxy in 24 hours,\\ with a peak flux density $S_{\rm peak} = 4$~mJy, a velocity width of\\ 150~km/s, and a prior photometric-redshift uncertainty of $\Delta z = 1$. \\
            $\bullet$ To detect \cii{} 158 $\mu$m or put a meaningful constraint on its\\ flux in 24 hours. \\
            $\bullet$ Forty (20) spectroscopically-identified galaxies for goal (baseline).
        \end{tabular} \\
\hline

\multirow{12}{*}{\begin{tabular}{l}Level 3:\\ Instrument\\ requirements\end{tabular}}
    &  \begin{tabular}{l}Technical\\ parameters\end{tabular}
        &  \begin{tabular}{l}
            $\bullet$ Tunable frequency range. \\
            $\bullet$ Instantaneous on-sky frequency coverage. \\
            $\bullet$ Spectral resolution. \\
            $\bullet$ Sensitivity.\\
            \end{tabular}\\
    \cline{2-3}
    &  \begin{tabular}{l}Technical\\ requirements\end{tabular}
        &  \begin{tabular}{l}
            $\bullet$ Tunable frequency range: 120--360~GHz with no gap (except for\\ the telluric water vapor lines at 183 and 325~GHz). \\
            $\bullet$ Instantaneous on-sky frequency coverage: $> 30$~GHz ($\Delta z > 1$ for\\ \oiii{} at $z = 10$) with as small a gap as possible. \\
            $\bullet$ Spectral resolution: 20~MHz ($\Delta V = 50$ km/s at 120~GHz,\\ 17~km/s at 350~GHz).\\
            $\bullet$ Sensitivity: Receiver noise temperature (single sideband) of 45~K\\ for 120--210~GHz and 75~K for 210--350~GHz, yielding system noise\\ temperature of $T_{\rm sys} \le 100$~K at 120~GHz, $\le 200$~K at 300~GHz,\\ $\le 300$~K at 350~GHz under a precipitable water vapor (PWV) of\\ 2~mm.
        \end{tabular}\\
\hline

\begin{tabular}{l}Level 4:\\ Mission\\ requirements\end{tabular}
    &  \multicolumn{2}{l}{
        \begin{tabular}{l}
            $\bullet$ PWV $\le$ 2~mm (270--350~GHz), $\le$ 3~mm (120--270~GHz).\\
            $\bullet$ Surface error $\le 75~\mu$m RMS ($> 270$~GHz), $\le 90~\mu$m RMS ($< 270$~GHz).\\
            $\bullet$ 10~hours per source (including overheads) times 40 (20) nights.
        \end{tabular}
        } \\
\hline
\end{tabular}
\end{table}

\begin{table}
\centering
\caption{FINER specifications.}
\label{tab:3}
\begin{tabular}{c|c|c}
\hline
\bf Item                &  \bf Baseline &  \bf Goal      \\
\hline
Frequency (Band 4+5)    &  125--210 GHz &  120--210 GHz  \\
\hline
Frequency (Band 6+7)    &  210--350 GHz &  210--360 GHz  \\
\hline
Beam size               &  \multicolumn{2}{c}{$7 \times ({\rm Frequency / 200~GHz})^{-1}$ arcsec}  \\
\hline
No.\ of polarization    &  \multicolumn{2}{c}{2}   \\
\hline
$T_{\rm RX}$ (Band 4+5) &  \multicolumn{2}{c}{45 K  ($>95\%$ of IF band)} \\
\hline
$T_{\rm RX}$ (Band 6+7) &  \multicolumn{2}{c}{75 K  ($>95\%$ of IF band)} \\
\hline
IF                      &  4--20.5 GHz  &  3--20.5 GHz  \\
\hline
Sideband separation     &  2SB, $>10$ dB &  2SB, $>25$ dB  \\
\hline
No.\ of basebands       &  8            &  16            \\
\hline
Bandwidth (wide)        &  81.92 GHz ($8\times 10.24$ GHz)&  163.84 GHz ($16\times 10.24$ GHz) \\
\hline
Spectral resolution (wide)&  \multicolumn{2}{c}{20 MHz} \\
\hline
Bandwidth (fine)        &  15.0~GHz ($6\times 2.5$ GHz) &  20.0~GHz ($8\times 2.5$ GHz) \\
\hline
Spectral resolution (fine)&  \multicolumn{2}{c}{88 kHz}  \\
\hline
Data dumping rate       &   \multicolumn{2}{c}{100 ms}  \\
\hline
\end{tabular}
\end{table}

\subsection{System}

\begin{figure}
    \centering
    \includegraphics[width=1.3\linewidth, angle=90]{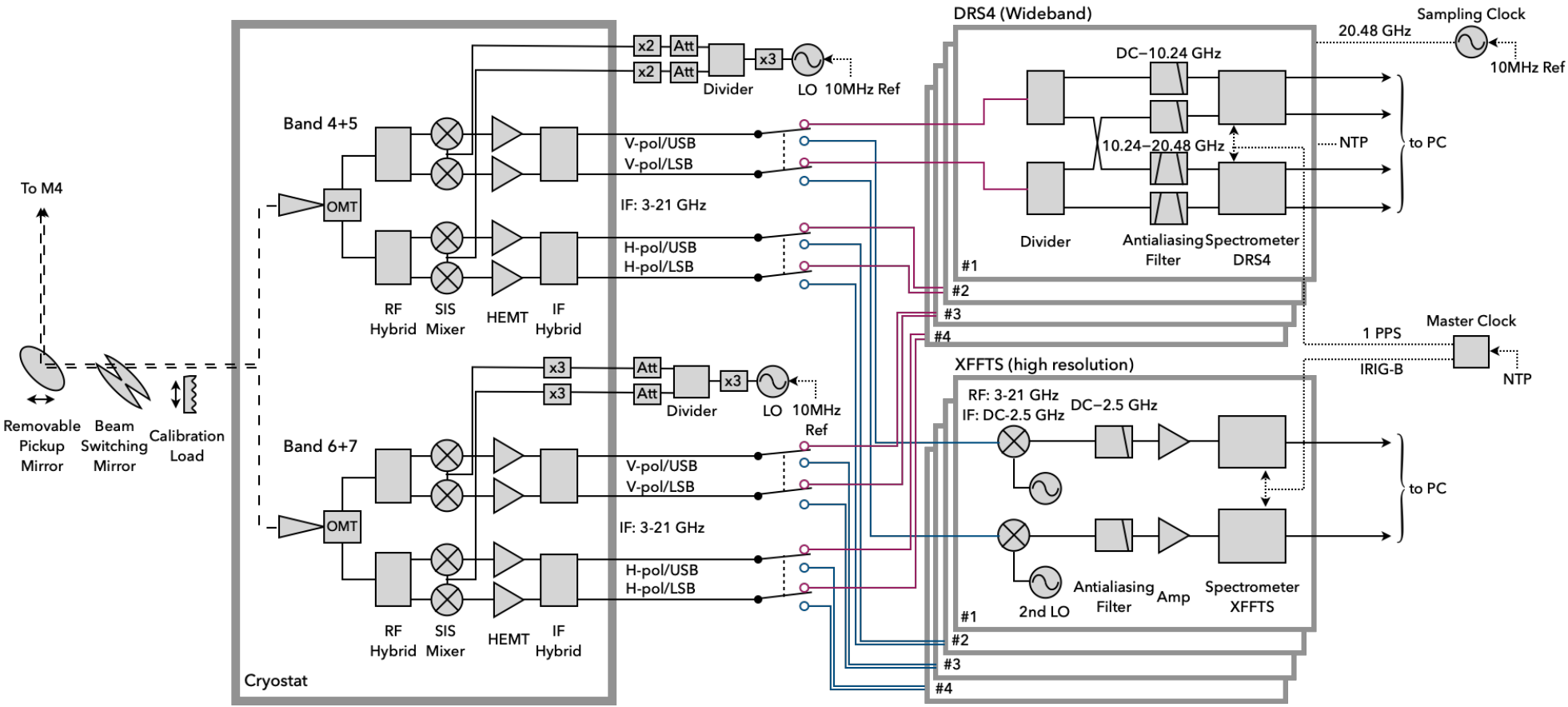}
    \caption{The system block diagram of FINER.}
    \label{fig:11}
\end{figure}

The system block diagram is shown in Figure~\ref{fig:11}. The system comprises the following subsystems; optics, frontend receiver, backend spectrometer, computers and software, which are described in the following sub-sections.

\subsubsection{Optics subsystem}
Warm optics include a pickup mirror (M5) and a chain of focusing mirror(s), a beam switching mechanism called the fast sky chopper (optional), and a load calibration device. The detailed optical parameters and layout are to be fixed (as of the time of writing). The pickup mirror is placed on a motorized linear slider and can be controlled remotely. 
The fast sky chopper is a set of removable mirrors to change the optical path slightly and to throw the on-sky beam position by $120''$ at a frequency of $\sim1$--2 Hz, which allows fast beam-switching observations. The number of available bands is limited to 1 when using the fast sky chopper due to the limitation of optical configuration.

\subsubsection{Front-end subsystem}
The receiver is coupled with the optics through corrugated feed horns. FINER uses the SIS receiver technology with wide RF and IF\cite{Kojima17, Kojima20}, which has been developed for ALMA Wideband Sensitivity Upgrade (WSU)\cite{Carpenter23}. Band 4+5 and 6+7 components are accommodated by a single cryostat, which is cooled with a two-staged GM 4-K cryocooler (RDK-415DP, Sumitomo Heavy Industry Inc.). Both bands are available simultaneously with the on-sky beam separation of $60''$. The observed frequencies of both bands are set independently. The receiver detects two polarizations in both upper and lower sidebands (USB and LSB) separately, yielding 4 IF outputs in each band. Note that it does not offer Stokes parameters other than intensity (i.e., Stokes-$I$).

\subsubsection{Back-end subsystem}
\begin{figure}
    \centering
    \includegraphics[width=0.5\linewidth]{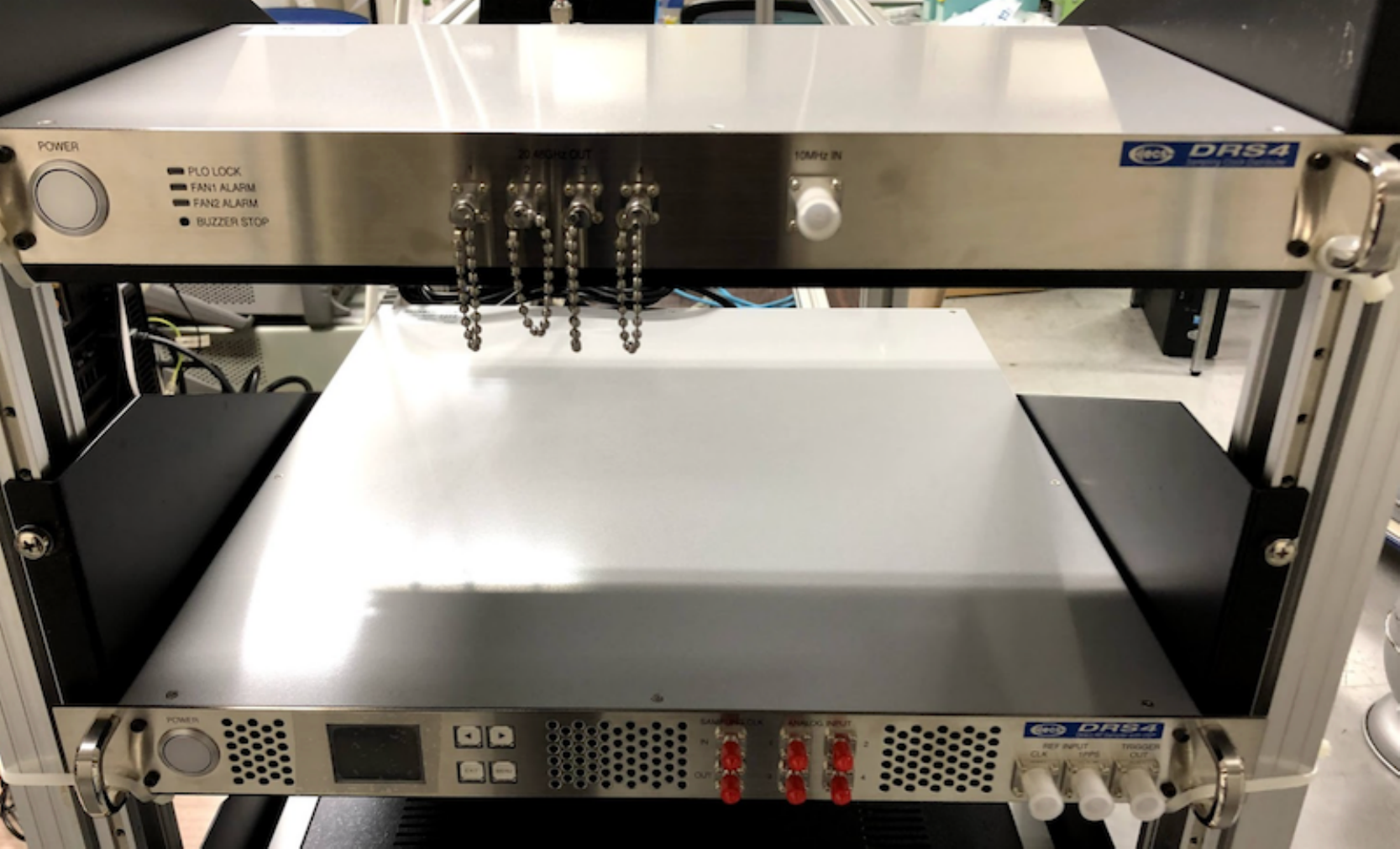}
    \caption{The digital spectrometer DRS4 with four 20.48 GSps samplers (bottom) and the 20.48 GHz reference signal generator (top).}
    \label{fig:13}
\end{figure}

The IF signals are sent to the spectrometers. FINER offers two types of spectrometers; wideband and high-resolution modes realized with DRS4 (Elecs Inc., Figure~\ref{fig:13}, Hagimoto et al., 2024 in this volume of Proc.\ SPIE) and XFFTS (RPG Inc.)\cite{Klein2012}, respectively. Users can select one of the spectrometer modes.

DRS4 directly samples the signal at 20.48~GSps with a good analog transmission up to 20.48~GHz. This offers the capability of spectroscopy in the 10.24--20.48~GHz band with no secondary down-converters, in addition to the baseband (DC to 10.24~GHz). DRS4 is an FX-type spectrometer with an optional output of cross-correlation for detailed calibration such as digital sideband separation correction. The input signal is digitized with a 3~bit 20.48~GSps analog-to-digital converter and is Fourier-transformed with one of window functions, rectangular, Hann, and Hamming, and then is squared and integrated to produce a 512 channel power spectrum at a dumping rate of 100~ms, while it does not offer a polyphase filterbank capability due to a limitation of the computational resource. It also offers a digital sideband separation capability \cite{Finger15, Rodriguez18}, to compensate for the imbalance of complex gain (i.e., amplitude and phase) between the sidebands and to remove the image sideband signal digitally (under development). 

XFFTS is an off-the-shelf spectrometer with high spectral resolution (88~kHz, corresponding to $\approx 0.13$~km~s$^{-1}$ at 200~GHz) and a 2.5-GHz bandwidth.\cite{Klein2012} Six boards of XFFTS are operational for B4R, and we will use them for FINER.

\subsubsection{Computer and software}

A control computer at the LMT backend cabin controls receiver tuning including local oscillator (LO) frequency setup and LO power optimization, spectrometer mode selection, pickup mirror drive, beam switching, and load calibration. The spectrometer outputs are recorded with NTP-synced timestamps on a data acquisition computer in the LMT backend cabin at a dumping rate of 100~ms. FINER is synchronized with LMT's telescope control system (TCS) through TCP/IP or UDP connection and does not require any hardware reference clocks from the TCS.

\subsection{Hardware Configuration}

\begin{figure}
    \centering
    \includegraphics[width=0.42\linewidth]{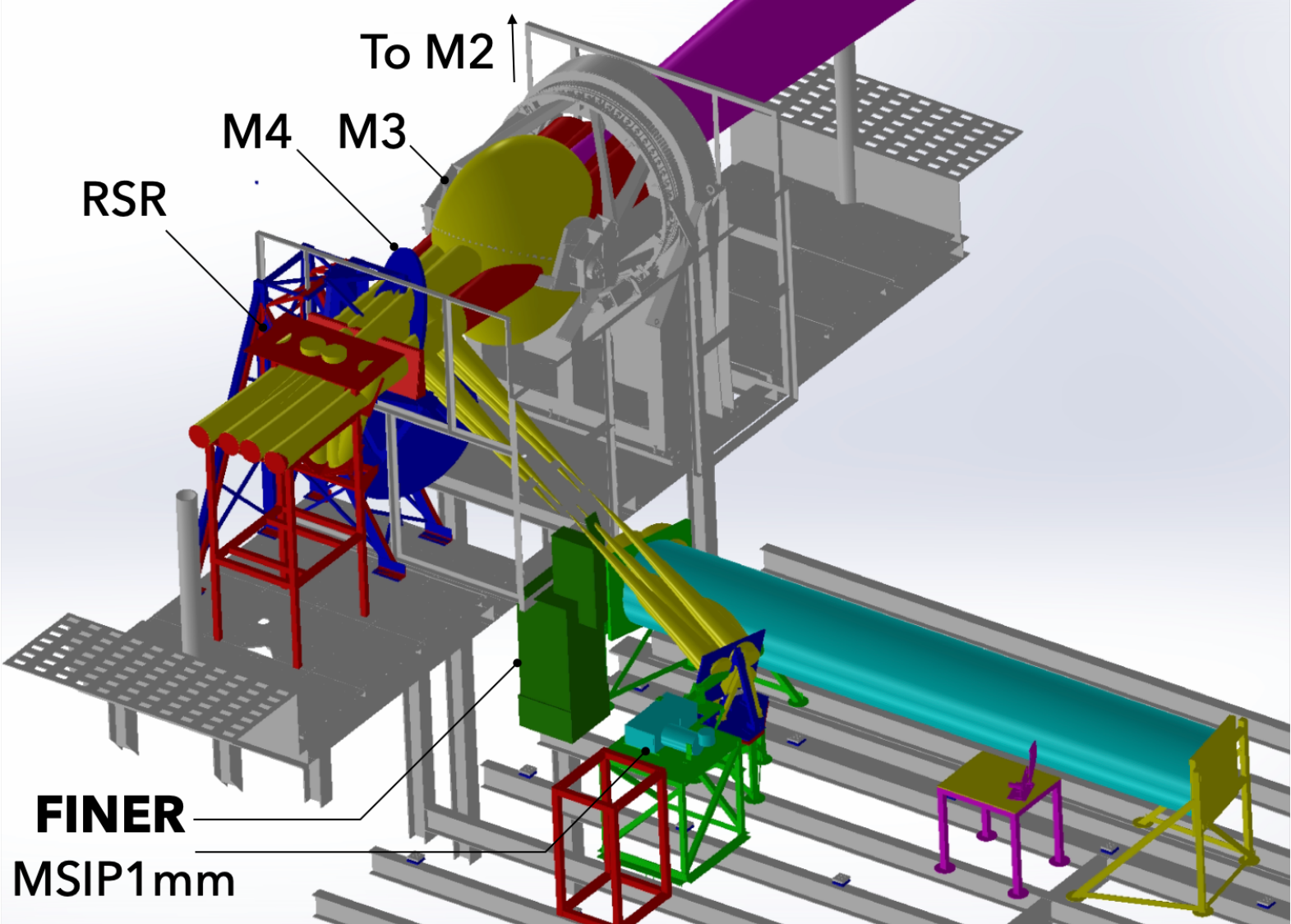}
    \includegraphics[width=0.45\linewidth]{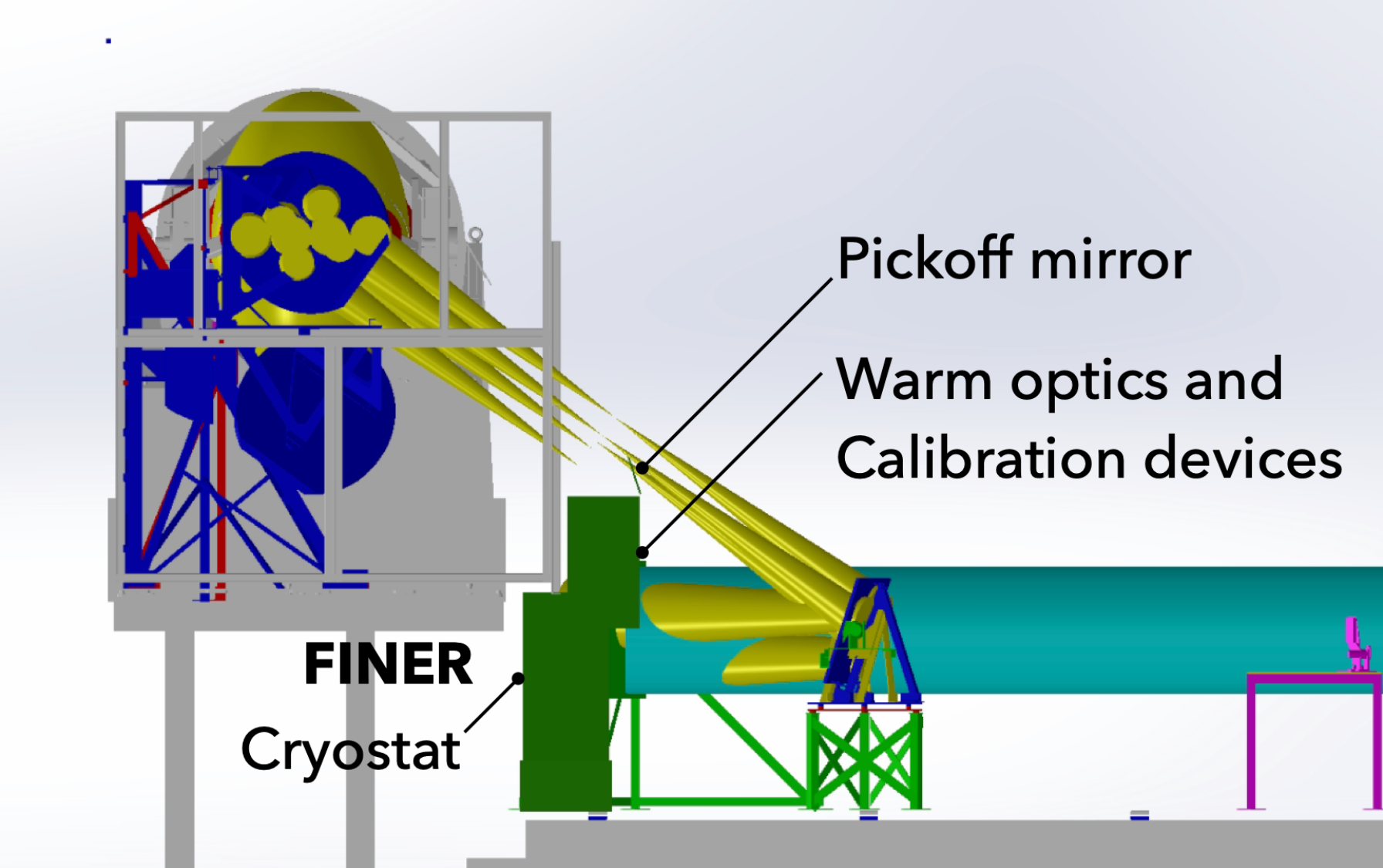}
    \caption{The configuration of the FINER receiver and warm optics in the LMT receiver cabin. The left and right panels show the CAD layout in bird’s eye view and from the direction parallel to the elevation axis, respectively. FINER is indicated by a green box. The beam ray from the fourth mirror (M4) to the fifth mirror (M5) is represented as the yellow bars, part of which is picked off by the FINER warm coupling optics. The location of FINER is similar to that of B4R.}
    \label{fig:14}
\end{figure}

\begin{figure}
    \centering
    \includegraphics[width=0.35\linewidth]{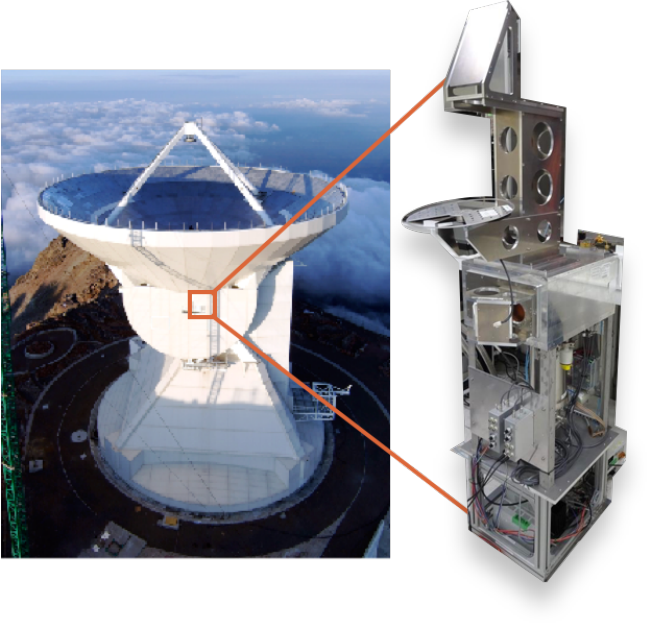}
    \caption{The B4R receiver system.}
    \label{fig:15}
\end{figure}

The proposed optics layout is shown in Figure~\ref{fig:14}. As we intend to replace the existing B4R receiver\footnote{The B4R (band 4 receiver) is the 2-mm receiver system which consists of a single-beam two-polarization SIS receiver operating at 125--163~GHz and four XFFTS spectrometers. B4R is based on the ALMA band-4 2SB mixer technology \cite{Asayama2014}. See \texttt{http://lmtgtm.org/telescope/instrumentation/instruments/b4r/?lang=en} for details.} (PI: Kawabe, Figure \ref{fig:15}) with FINER, the layout is almost the same as B4R; The ray from M4 to M5 (yellow beam toward SEQUOIA) is bent by a removable pickup mirror. The bent ray is directly led into the FINER cryostat and is coupled with the two feed horns of Band 4+5 and 6+7. FINER has a beam-switching mechanism in between the pickup mirror and the cryostat, as well as a chopper wheel for the standard single-load calibration.

The cryostat is placed at the position of B4R in the LMT receiver cabin. We will replace B4R's cryostat and cryocooler with new ones, while the helium hoses and chiller of B4R will be used for FINER. The LO system and digitizers are placed alongside. The IF signal is sent to the spectrometers in the backend cabin downstairs.

\subsection{Data Analysis Software}

The observed data will be provided in the \texttt{MeasurementSet} (MS) format of \textsc{casa}\cite{CASA22}, in which time series spectra taken on- and off-source and for load calibration are contained along with telescope boresight positions, ancillary environmental records and metadata.

Users will receive MSs as initial datasets and will use \textsc{casa} for fundamental data reduction including intensity and bandpass calibration, flagging, velocity correction, spectrum making, and map/cube making. Further analysis such as spectral fit can also be made with the single-dish tasks of \textsc{casa}.
A new data-scientific noise-removal method demonstrated using the B4R receiver\cite{Taniguchi2021}, which reduces the baseline ripples and improves the observed sensitivity by a factor of $\sqrt{2}$, will also be provided by the development team.

\section{DEVELOPMENT PLAN}\label{sect:DevPlan}

The development is led mainly by Nagoya University, the University of Electro-Communications (UEC), Advanced Technology Center (ATC) at NAOJ, Kitami Institute of Technology, the University of Tokyo, and Keio University in collaboration with the LMT project. Development, commissioning, and early science operations of FINER are planed in 2022--2027. We plan to deliver the instrument to LMT and to start commissioning and science operations in 2025/26.


At the initial phase FINER will begin with the `Baseline' specifications (Table~\ref{tab:3}) and will aim for the `Goal' specifications when technologies and resources are ready. As the SIS mixer technology is likely being improved after FINER gets operational, we may upgrade the SIS mixer device if this is the case. The back-end spectrometers are also added to enhance the capability. The data reduction pipeline will be improved by implementing a new algorithm for noise removal.\cite{Taniguchi2021}
The FINER cryostat and warm optics are designed so that another additional single-beam receiver, such as a 3--4~mm band receiver, may be accommodated in the future, without any conflict with the 120--360~GHz band (see the FINER+ parameter space in Figure~\ref{fig:5}).

\acknowledgments 
 
This study is financially supported by JSPS KAKENHI (No.\ 22H04939, 23K20035), NAOJ Joint Development Research Programs (No.\ 1901-0101, 2001-0104), and the Grant for Joint Research Program of the Institute of Low Temperature Science, Hokkaido University (23G037 and 24G036)

\bibliography{report} 
\bibliographystyle{spiebib} 

\end{document}